\documentclass[11pt]{article}
\usepackage{authblk}
\usepackage{graphicx}
\usepackage{amsmath}
\usepackage{amsfonts}
\usepackage{amssymb}
\usepackage{fontenc}
\usepackage{mdframed}

\newlength{\headwidth}
\newlength{\dinwidth}
\newlength{\dinmargin}
\newlength{\medwwidth}
\newlength{\hnumwidth}

\begin{document}
\title{\bf How quarks made their entrance into our worldview \\
{\large via cosmic rays and proton accelerators}}
\author{Jos Engelen}
\affil{NIKHEF, Science Park 105, 
1098 XG Amsterdam, The Netherlands}
\maketitle
\section*{1964: a new layer in the structure of matter}
Sixty years ago Gell-Mann introduced, with a striking sense of caution, a new fundamental layer in the structure of matter. He christened the associated particles `quarks', a name borrowed from the psychedelic flood of words of James Joyce's novel Finnegans Wake. Independently and simultaneously, Zweig arrived at similar ideas. We outline the circumstances that led to the introduction of quarks.
\section*{A new interaction, a new quantum number: isospin}
With the discovery of the neutron (Chadwick, 1932) a new field of study began: the exploration of the `strong interactions' that bind proton (p) and neutron (n) into atomic nuclei. In a first attempt to describe these new nuclear forces (\cite{Heisenberg}, also in 1932), Heisenberg introduced a new quantum number, which he did not give a name, but which later came to be called isospin. Isospin provided a convenient way for Heisenberg to distinguish the interactions between nucleons (a collective name for protons and neutrons) in an atomic nucleus (pn, pp, nn) and provided a compact notation. It was only much later that the new quantum number proved more generally relevant when more strongly interacting particles were discoverded. Particles that experience this interaction are called `hadrons' (hadros=strong) which occur in two categories, baryons (barys=heavy) such as proton and neutron (with half integer spin: fermions) and mesons (mesons=middle) (with integer spin: bosons) such as the yet-to-be-introduced pion.

For Heisenberg, proton and neutron were two states of the same particle in a new abstract space, in analogy to the two spin states of a spin 1/2 particle in ordinary space. The attractive force between proton and neutron was caused by a particle jumping back and forth between the two. This, according to Heisenberg, led to `Platzwechsel' of proton and neutron and back again, so that proton and neutron stayed close together. 

To illustrate this, Heisenberg uses the $\sf H_2^+$ ion where the electron jumps back and forth between protons. Also for the proton-neutron interaction, the electron was the only known particle that could perform this role, but then conservation of (ordinary) spin would be in question: proton, neutron and electron all three have spin 1/2.
An important next step was taken by Yukawa (in 1935, \cite{Yukawa}). He argued that a new field, with short range but strong coupling, was responsible for the force between proton and neutron and that the corresponding field quantum was the particle causing `Platzwechsel'. A rough estimate of its mass, leading to the short range, came out to be two hundred times the electron mass. Such a particle was not known and Yukawa reasoned that it was unobservable and did not occur as a free particle. Physicists of the time were very reluctant to predict new particles! It was not until 1947 that the first particle was discovered that exhibited the strong interaction in addition to proton and neutron: the pion or $\sf \pi$-meson \cite{pion1947}. The pion was discovered by analyzing the tracks left by cosmic rays in photographic emulsions. The name meson was chosen to indicate that the particle in question had a mass somewhere between the light electron (a lepton, leptos=small) and the heavy proton. The mass of the pion was about 300 electron masses. The pion carried charge, assumed to be the elementary charge, and interacted with atomic nuclei. The tracks were recorded using photographic emulsions that detected ionizing cosmic rays. The discovery of an electrically neutral pion followed a few years later at the first proton accelerators.

Yukawa's original idea, description of the proton-neutron interaction by a charged pion jumping back and forth was extended to strong p-p and n-n interactions via exchange of a neutral pion. This led to the picture of charge independence of the strong interactions. In addition to proton and neutron, the pion was also assigned isospin and with value 1 so that the three projections 1,0,-1 (remember again the analogy with ordinary spin) corresponded to the three charge states of the pion. Charge independence of the strong interaction is then synonymous with isospin symmetry and thus with conservation of isospin.

The three pions, $\sf \pi^+,\pi^0,\pi^-$ and proton and neutron form isospin multiplets (a triplet and a doublet, respectively). The particles within a multiplet have almost but not exactly equal mass which indicates that isospin symmetry in nature is not perfect, but somewhat `broken'.

With the first proton accelerators that came into operation in the early `50s of the $\sf 20^{th}$ century, it was possible to create charged pion beams and then scatter them off protons (liquid hydrogen) - this is how Fermi (gifted theorist as well as experimentalist) conducted experiments at the cyclosynchrotron of the University of Chicago. There the first `resonance' was found: the number of pion-proton collisions (the total cross section) showed a strong maximun at a beam energy corresponding to 1230 MeV in the center of mass system. This maximum had a width at half height of 100 MeV.
This `resonance' is a particle with a mass of 1230 MeV and a lifetime of 1/100 MeV$\sf^{-1}$ = 10$\sf^{-23}$ s. The particle was named $\sf \Delta$. The data are consistent with isospin 3/2, i.e., the $\sf \Delta$ occurs in four charge states (++,+,0,-). The isospin formalism is summarized in the box.
At higher-energy proton accelerators in the `50s and `60s (CERN's proton synchrotron came into operation in 1959) many more resonances were found, belonging both to the baryons and mesons. We will return to this later. First we introduce a second new quantum number discovered in the context of the strongly interacting particles, such as the pion, found in cosmic rays.
\\
\begin{figure}[t]
   \begin{mdframed}
  % \begin{footnotesize}
  { \bf The isospin formalism} \\ \\
   $\sf |\pi^+>=|I,I_3>=|1,1>$, $\sf |p>=|1/2,1/2>$, then $\sf |\pi^+p>=|3/2,3/2>.$
   \\ The $\sf \Delta^{++}$ resonance was discovered in the reaction\\
     $\sf \pi^+p\rightarrow \Delta^{++}\rightarrow \pi^+p$ so has $\sf |I,I_3>=|3/2,3/2>$.	
   Similarly, $\sf \pi^-p\rightarrow \Delta^{0}\rightarrow \pi^-p$.
   But here, there is a second possibility:
   \\	$\sf \pi^-p\rightarrow \Delta^{0}\rightarrow \pi^0n$. \\
    In terms of isospin, we write:\\	$\sf |\Delta^{0}>=| 3/2,-1/2>\rightarrow  A |1,-1;1/2,1/2>+B|1,0;1/2,-1/2>$.	\\
    Here A and B are the Clebsch-Gordan coefficients, known from the quantum mechanics of angular momentum. In more mathematical terms, they are the coefficients that appear in the direct sum of the decomposition of the product of two irreducible representations of the symmetry group of angular momentum: SU(2). In our example, $\sf A=\sqrt{1/3}$ and $\sf B=\sqrt{2/3}$, and this means that conservation of isospin implies that the $\sf \Delta^0$ decays to
   $\sf \pi^-p$ in 1/3 of the cases and in 2/3 of the cases to $\sf \pi^0n$,
   which has been experimentally verified.	\\																	\end{mdframed}
\end{figure}										
\section*{Another new quantum number: strangeness}
In addition to charged pions, cosmic rays had other new particles in store, produced in collisions of energetic ($>$1 GeV) protons.

\begin{figure}[h!]
\centerline{\includegraphics[width=14cm]{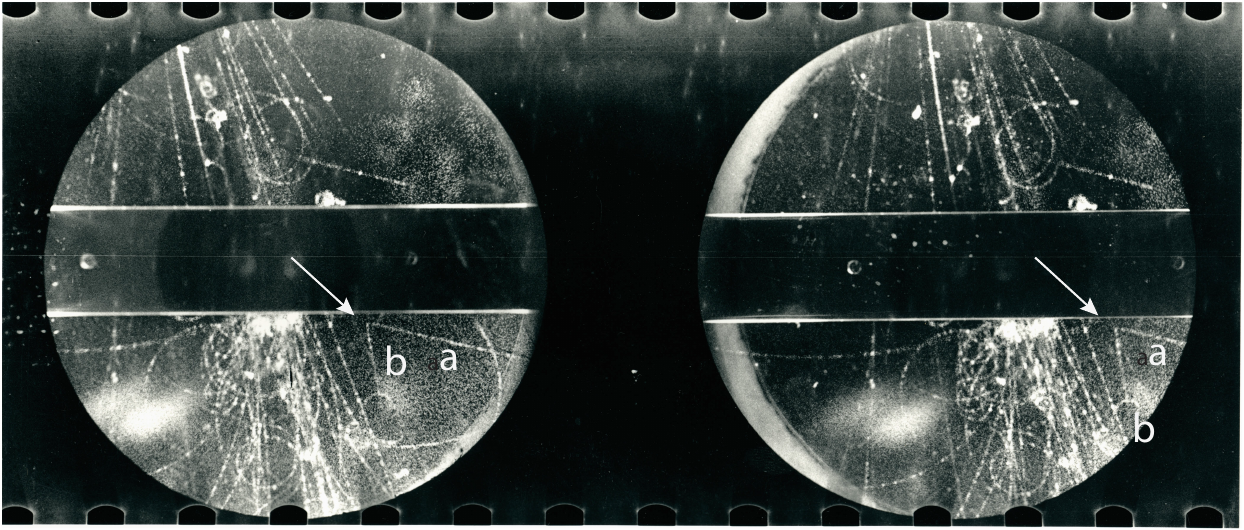}}
\caption{Stereo cloud chamber picture of $\sf K^0$ decay. The broad band in the center is a lead sheet 3 cm thick to recognize `penetrating particles'. The $\sf K^0$ decays into two tracks a and b, which seem to come out of nowhere: the white arrow points to the point where the neutral kaon decays. See \cite{Rochester}. Image courtesy Robin Marshall.}
\label{K0}
\end{figure}

Besides photographic emulsions, delivered by balloons high into the atmosphere, also cloud chambers on high mountain tops, such as the Pic du Midi (2885 m) in the French Pyrenees, were used to detect cosmic rays, or particles produced by cosmic rays. Initially, cloud chambers were used only in laboratories at sea level - transporting to and operating the equipment, especially the heavy magnets, on high mountain tops was quite an undertaking. In Fig. \ref{K0} we see a cloud chamber picture, taken in Manchester, showing `out of nowhere' two tracks of charged particles coming from one and the same point.

After careful analysis, no other conclusion was possible than that this was a new, uncharged, particle that decayed into two charged pions. Apart from the `forks' (`V particles') of Fig. \ref{K0}, tracks with a distinct kink were also found, interpreted as the decay of a newly charged particle into one charged and one neutral pion. The mass of these new particles was about 1000 times the electron mass. They were referred to as `kaons' and occurred in three charge states:  $\sf K^0,K^+$ and $\sf K^-$. These particles were initially a great mystery. They were produced abundantly in high-energy collisions of cosmic protons which indicated the same strong interaction that underlies the production of pions. The kaons decayed to two pions. However, the lifetime was of the order of $\sf 10^{-10}$ s, at least eleven orders of magnitude longer than might be expected based on the strong interaction. Pioneering contributions to the solution of the puzzle were made by Abraham Pais, emigrated from the Netherlands immediately after World War II, working at Princeton. In essence, he proposed that the kaons carried a new `scalar' quantum number that was conserved in strong interactions. The kaons were therefore produced in kaon-antikaon pairs (`associate production') where the sum of this new quantum number was zero. For each individual kaon it had a value +1 or -1 so that decay according to the strong interaction in pions, for which this quantum number was 0, was forbidden. The decay $\sf K^0\rightarrow \pi^+\pi^-$, as shown in the historical picture of Fig. \ref{K0}, thus had to proceed according to a new weak interaction, to be further investigated, which gave rise to a much larger lifetime. This new quantum number was given the name strangeness.

\begin{figure}[h!]
\centerline{\includegraphics[height=10cm]{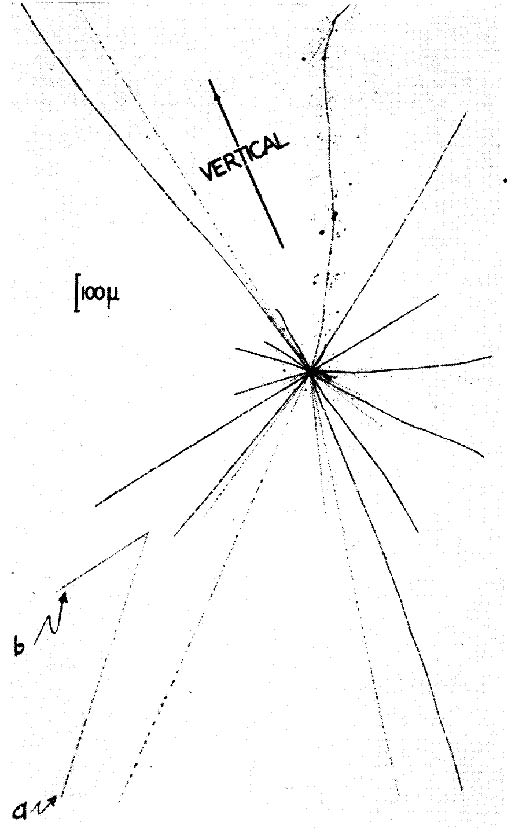}}
\caption{Facsimile of registration in photographic emulsion of a cosmic ray (proton) `blowing up' a heavy nucleus (silver or bromine). In addition and independently, two tracks are visible that appear to come out of nowhere. Based on the grain density (`ionization'), trace b is identified as a proton. Here, we see $\sf \Lambda \rightarrow p\pi^-$. See \cite{Hopper}.}
\label{Lambda}
\end{figure}

In 1950, the observation of a `V particle' like the kaon but with a proton as one of the decay particles was reported from Australia \cite{Hopper}. Fig. \ref{Lambda} shows the picture on which the $\sf \Lambda$ decay is recorded. This is a photographic emulsion exposed to cosmic rays at an altitude of 70,000 feet. The typical `star' caused by an incident proton is not what is at issue. The out of nowhere tracks a and b, where the grain density of track b indicates a proton form the basis of the discovery of the $\sf \Lambda$ baryon. The center of the `star' is not in the plane of a and b, so the $\sf \Lambda$ particle is not related to the nuclear reaction causing the star.

So the $\sf \Lambda$ particle decays, like $\sf K^0$, into particles that know the strong interaction, $\sf \Lambda\rightarrow p\pi^-$, but has a large lifetime indicating a `weak' decay. Thus, $\sf \Lambda$ also carries strangeness.

\section*{More new particles: arrangement in multiplets - a new symmetry}

\begin{figure}[h!]
\centerline{\includegraphics[width=12cm]{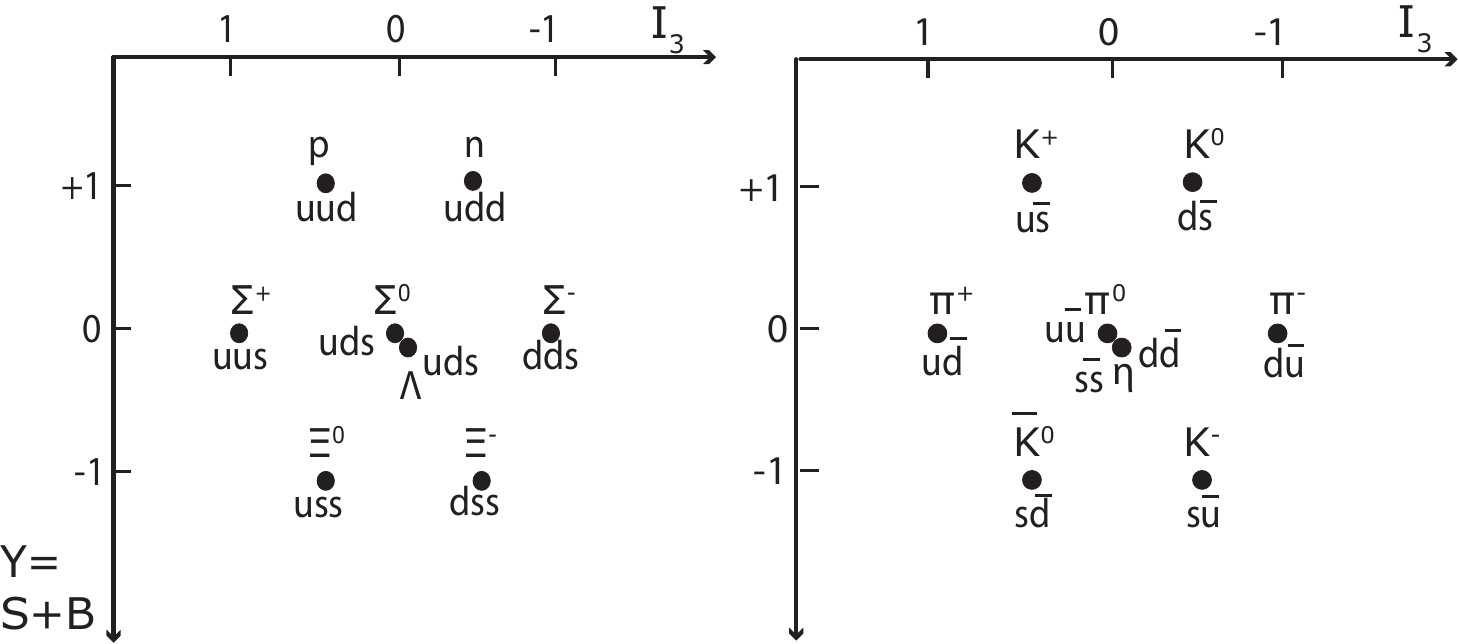}}
\caption{
Octet of spin 1/2 baryons and of spin 0 mesons. The composition in terms of quarks (see below) is also indicated.}
\label{Octets}
\end{figure}

Observations using photographic emulsions and cloud chambers of cosmic rays and particles produced in collisions in the high atmosphere led to the discovery of new particles that interacted `strongly'. Supplemented by the results of the proton accelerators that came into operation from the early `50s of the 20$\sf^e$ century, such as the synchrocyclotron of the University of Chicago mentioned earlier, these particles showed an arrangement in isospin multiplets and, if `strangeness' was also included, super-multiplets. We illustrate this in Fig. \ref{Octets} where the octets of baryons and mesons are shown.

In 1961, Gell-Mann published `the eightfold way' \cite{MGM61} following the observed octets (a name referring to Buddhism's noble eightfold path to enlightenment). Independently, Ne'eman arrived at a similar analysis \cite{YN61}. The idea was simple: just as isospin-multiplets imply invariance of the strong interactions under isospin rotations (to be represented as matrices belonging to the group SU(2)), so the observed super-multiplets imply invariance of the strong interactions under isospin/strangeness rotations to be represented as matrices belonging to the group SU(3). Put slightly more carefully: isospin multiplets correspond to irreducible representations of SU(2); isospin/strangeness super-multiplets correspond to irreducible representations of SU(3) \footnote{The group of unitary 3x3 matrices with determinant 1}.

Initially there was no obvious reason for choosing SU(3), but we now know that reason. In `the eightfold way' the simplest, namely the three-dimensional representation of SU(3) was ignored. We quote the reason for this here: the young Nijmegen professor De Swart wrote in 1963 \cite{deswart}, one year before the introduction of quarks {\it `the lowest nontrivial IR in the
octet model, which is physically possible (i.e., has integer quantum numbers for the hypercharge), is
the IR \{8\}.'} (IR stands for `irreducible representation'. `Hypercharge' can be read as baryon number or electric charge.) In his much cited article following the octet classification of Gell-Mann and Ne'eman, he thus misses the opportunity for a great discovery, which was contained in taking the `impossible'  three-dimensional representations seriously.
%\{3\} en \{$ \overline{3}$\}.  
This is exactly what Gell-Mann \cite{GellMann} and independently and almost simultaneously Zweig \cite{Zweig} did. Thus they postulated a new layer in the structure of matter consisting of three particles with `fractional' charge and fractional baryon number. Zweig called these particles `aces' and Gell-Mann `quarks', the name that stuck. Quarks were highly controversial upon their introduction and also in the years that followed. Zweig did not get his manuscript published; it always remained a CERN preprint. And Gell-Mann was very careful to play down his claims.  He writes at the end of his article:
{\it It is fun to speculate about the way quarks would
behave if they were physical particles of finite mass
(instead of purely mathematical entities as they
would be in the limit of infinite mass).} 
He then reasons that the lightest quark must be stable and might be found in the Earth's crust. The last sentence of the article, which by the way is only two pages long, contains a `disclaimer', a kind of excuse for bringing up quarks: {\it  A search
for stable quarks of charge -1/3 or +2/3 ... at the highest
energy accelerators would help to reassure us of
the {\bf  non-existence} of real quarks.} (Emphasis added).
\begin{figure}[h!]
\centerline{\includegraphics[width=14cm]{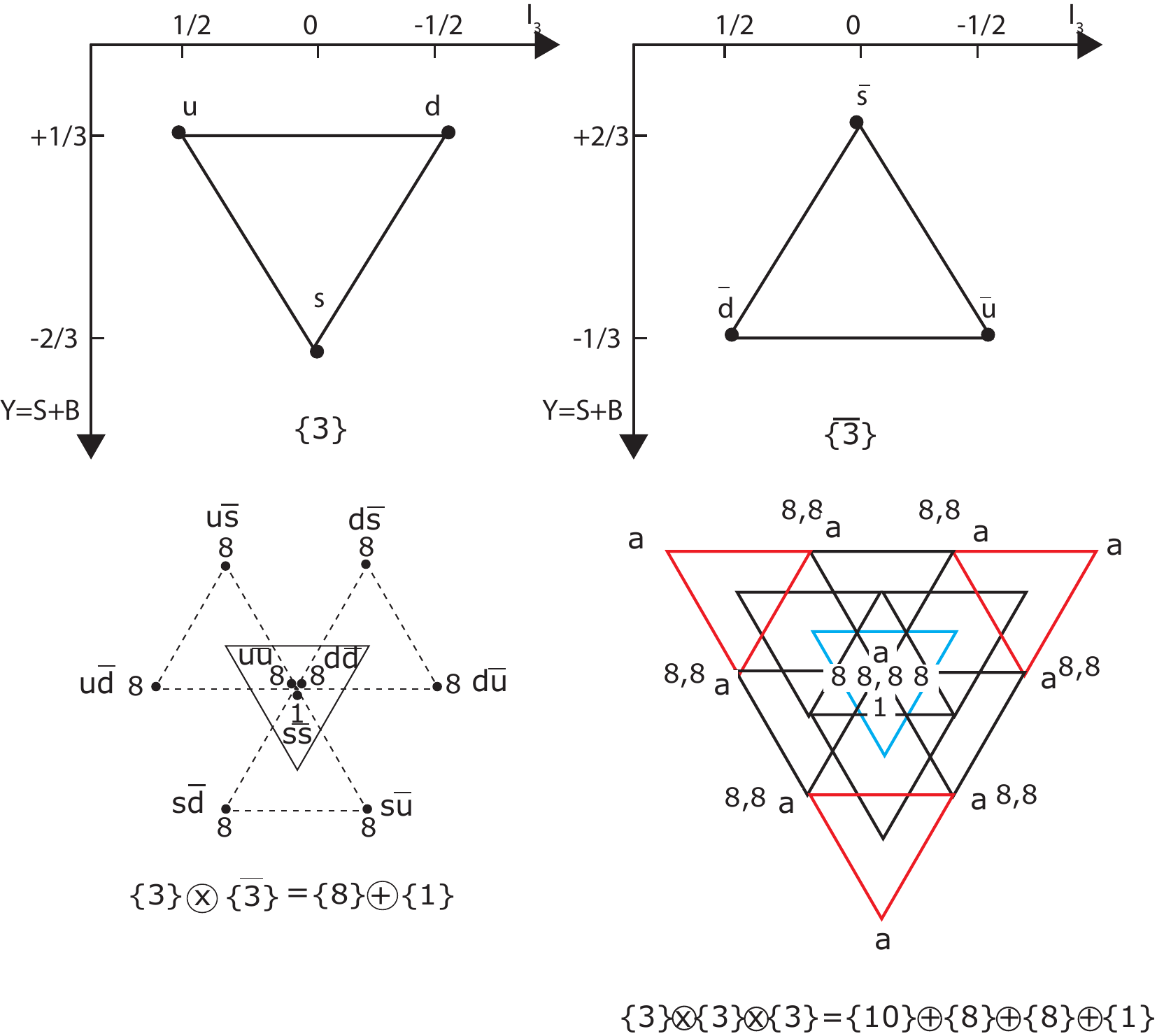}}
\caption{
Triplets of quarks and anti-quarks. Bottom left octet \{8\} and singlet \{1\} of quark-antiquark combinations, mesons. Bottom right an illustration of the representations that can be constructed by combining three quarks: a decuplet (a), two octets (8), a singlet (1).}
\label{quarks}
\end{figure}

In Fig. \ref{quarks} we illustrate how mesons and baryons can be constructed from the fundamental triplets (the quarks).

\section*{Quarks as `static' building blocks versus quarks as dynamical particles}

The description of hadrons as composed of three quarks (baryons) or quark-antiquark pairs (mesons) is in good agreement with observations. The quarks as they follow from the models of Gell-Mann and Zweig occur in three `flavors': u (up), d (down), s (strange), also called `SU(3) of flavor'. The symmetry under rotations in the space spanned by (u,d,s) is not perfect because the particles within the various (super-)multiplets do not have exactly equal masses, but apart from that, the observed hadrons fit the scheme very nicely. The baryon decuplet was initially incomplete, a particle was missing. Its discovery, the $\Omega^-$ baryon, consisting of three s quarks, was a spectacular confirmation of the quark model.

Still, the quark model was not satisfactory. For example, the $\Omega^-$ baryon has spin 3/2 and is therefore composed of three s quarks with the spins parallel. However, according to the Pauli principle, this is forbidden because a state of multiple fermions must be anti-symmetric under exchange of two fermions. 

Furthermore, there was no success in constructing a field theory according to the principles of local gauge invariance (under SU(3) transformations of `flavor') as set forth by Yang and Mills in a 1954 visionary paper. This was the case until until 1973. The same Gell-Mann, along with Fritzsch and Leutwyler, published the theory of quarks and gluons now known as QCD (quantum chromodynamics) \cite{FGL}. QCD is invariant under local SU(3) gauge transformations of a fundamental triplet not of `flavor' but of `color': each quark can carry one of three `color charges'.

In 1966 in Menlo Park, California, the two-mile linear electron accelerator at the Stanford Linear Accelerator Center (SLAC) came into operation that could accelerate electrons to an energy of 20 GeV. Experiments that studied electron-proton scattering in this new energy domain showed surprising results. In the 1950s, R. Hofstadter had used electron beams with energies of several hundred MeVs, in part to demonstrate that protons were not point particles. The measurements showed that the scattering took place off a charge sphere with a radius of about 1 fm ($\sf 10^{-15}m$). In 1969, the first results of SLAC at much higher energy were published \cite{dis}. The scattering was found not to take place exclusively off a diffuse charge sphere, `events' were also observed in which the electrons were scattered at a large angle, and the momentum transfer associated with this indicated scattering off an object much smaller than the proton. This marked the beginning of the description of the proton (and hadrons more generally) as a dynamical system of bound quarks, antiquarks and gluons obeying the laws of QCD. At short distances, much smaller than the proton, the coupling constant is small (`asymptotic freedom') so that quantitative calculations are possible. At larger distances, however, the coupling constant increases and leads to `confinement' of quarks into hadrons, a mechanism not yet understood in detail. For an article on the development of QCD by one of the `main characters', please refer to \cite{thooft2}.

The three quarks underlying the original quark model turned out not to be the only ones found in nature. It is beyond the scope of this article but three more `quark flavors' were discovered: `charm' (c), `beauty' (b) and `top' (t). The discovery of charm quarks as building blocks of the $\sf J/\psi$ particle in 1974 confirmed, should there still have been any doubt, the physical nature of quarks as tangible particles. The discovery of b and t followed in 1977 and 1995, respectively.
There are reasons to believe that these six quarks complete the picture. All six quarks, by the way, fit quantitatively into QCD. They exhibit significant mass differences that are not yet understood.

With this we conclude the history of the introduction of quarks into our worldview. They are now established ingredients of the Standard Model. 

\subsection*{Acknowledgements}
I thank Karel Gaemers for a number of discussions on earlier versions of this article, which greatly influenced its final content. I also thank Piet Peters and Henk Tiecke for critically reading through this article.


\newpage

\begin{thebibliography}{99}

\bibitem{Heisenberg} Heisenberg, W. Über den Bau der Atomkerne. I. Z. Physik 77, 1–11 (1932).
\bibitem{Yukawa} Yukawa, H. (1935). On the Interaction of Elementary Particles. I. Proceedings of the Physico-Mathematical Society of Japan, 17, 48-57; Reprinted in: Progress of Theoretical Physics Supplement, Volume 1, January 1955, Pages 1–10,
\bibitem{pion1947} Lattes, C., Muirhead, H., Occhialini, G., Powell C.F., Processes Involving Charged Mesons. Nature 159, 694–697 (1947). https://doi.org/10.1038/159694a0
\bibitem{Rochester} G.D. Rochester and C. C. Butler, {\it Evidence for the existence of new unstable elementary particles}, Nature 160 (1947) 855-857. 
\bibitem{Hopper} Hopper, V. D. and Biswas, S. (1950). Evidence Concerning the Existence of the New Unstable Elementary Neutral Particle. Physical Review, 80(6), 1099–1100. doi:10.1103/physrev.80.1099 
\bibitem{MGM61}M. Gell Mann, The Eightfold Way: A Theory of Strong Interaction Symmetry, California Institute of Technology, Report CTSL-20 (1961).
\bibitem{YN61} Y. Ne'eman, Derivation of Strong Interactions from a Gauge Invariance,  Nuclear Physics 26 (1961) 222--229.
\bibitem{deswart} J.J. de Swart, The Octet model and its Clebsch-Gordan coefficients, Rev.Mod.Phys. 35 (1963) 916-939, Rev.Mod.Phys. 37 (1965) 326-326 (erratum)
\bibitem{GellMann} M. Gell-Mann, A Schematic Model of Baryons and Mesons, Phys.Lett. 8 (1964) 214-215.
\bibitem{Zweig} G. Zweig, An SU(3) Model for Strong Interaction Symmetry and its Breaking, CERN preprint 8419/TH.412 (21 February 1964).
\bibitem{FGL} H. Fritzsch, M. Gell-Mann and H. Leutwyler, Advantages of the Color Octet Gluon Picture, Phys. Lett. B 47, 365 (1973).
\bibitem{dis} E. D. Bloom, D. H. Coward, H. DeStaebler, J. Drees, G. Miller, L. W. Mo, R. E. Taylor, M. Breidenbach, J. I. Friedman, G. C. Hartmann, and H. W. Kendall
Phys. Rev. Lett. 23, 930 (20 October 1969).
\bibitem{thooft2} G. `t Hooft, When was Asymptotic Freedom discovered? or The Rehabilitation of Quantum Field Theory. arXiv:hep-th/9808154v2 

\end{thebibliography}
\end{document}